\documentclass[useAMS,usenatbib]{mn2e}  
\usepackage{epsfig}   
   
\def\eff{_{\rm eff}}         
\def\eq{equation }         
\def\eqs{equations }       
\def\fig{Fig.~}   
           
\def\nnb{\nonumber}         
\def\H{_{\rm H}}          
          
\def\ph{_{\rm PH}}          
\def\b{_{\rm b}}          
\def\der{{\rm d}}          
\def\MM{_{\rm MM}}          
\def\min{_{\rm min}}          
\def\max{_{\rm max}}          
\def\g{_{\rm g}}          
\def\v{_{\rm v}}          
\def\bar{_{\rm b}}   
          
\def\gal{_{\rm G}}

\def\m{_{\rm m}}          
\def\em{_{\rm em}}          
\def\p2{_{\rm p2}}          
\def\p{_{\rm P}}

\def\nbody{{$N$}-body }          
\def\delm{\Delta\m}          
\def\delc{\delta_{\rm c}}          
          
\def\acc{_{\rm acc}}          
          
\def\Edd{_{\rm Edd}}  
  
\def\BH{_{\rm BH}}          
   
\def\cos{}          
\def\Th{_{\rm Th}}          
\def\last{_{\rm last}}

\def\form{_{\rm form}}          
\def\eff{_{\rm eff}}          
\def\top{_{\rm top}}          
\def\refe{_{\rm c}}          
\def\qb{_{\rm rr}}           
\def\eg{{e.g.\ }}  
\def\ie{{i.e.\ }}  
   
\title[The Luminosity Function of Quasars]{Major Mergers of Haloes,   
Growth of Massive Black Holes and Evolving Luminosity Function of   
Quasars}   
        
\author[Hatziminaoglou et al.]{Evanthia Hatziminaoglou$^{1}$, Guy   
Mathez$^{2}$, Jos\'e-Mar\'\i a Solanes$^{3}$, Alberto Manrique$^{4}$   
\newauthor and Eduard Salvador-Sol\'e$^{4,5}$\\   
$^{1}$ESO, Karl-Schwarzschild-Strasse 2, D-85748 Garching, Germany\\       
$^{2}$Laboratoire d'Astrophysique, Observatoire Midi-Pyr\'en\'ees,         
14 rue Edouard Belin, F-31400 Toulouse, France\\   
$^{3}$Dpt. d'Enginyeria Inform\`atica i Matem\`atiques, Universitat         
Rovira i Virgili, Avda. Pa\"\i sos Catalans, 26, E-43007 Tarragona, Spain\\   
$^{4}$Dpt. d'Astronomia i Meteorologia, Universitat de Barcelona,         
Mart\'\i\ Franqu\`es 1, E-08028 Barcelona, Spain\\         
$^{5}$CER for Astrophysics, Particle Physics and Cosmology,  
Universitat de Barcelona, Spain}  
   
\begin{document}        
   
\maketitle   
   
\begin{abstract}    
\noindent We construct a physically motivated analytical model for the  
quasar luminosity function, based on the joint star formation and  
feeding of massive black holes suggested by the observed correlation  
between the black hole mass and the stellar mass of the hosting  
spheroids. The parallel growth of massive black holes and host  
galaxies is assumed to be triggered by major mergers of haloes. The  
halo major merger rate is computed in the frame of the extended  
Press-Schechter model. The evolution of black holes on cosmological  
timescales is achieved by the integration of the governing set of  
differential equations, established from a few reasonable assumptions  
that account for the distinct (Eddington-limited or supply-limited)  
accretion regimes. Finally, the typical lightcurves of the reactivated  
quasars are obtained under the assumption that, in such accretion  
episodes, the fall of matter onto the black hole is achieved in a  
self-regulated stationary way. The predicted quasar luminosity  
function is compared to the luminosity functions of the 2dF QSO sample  
and other, higher redshift data. We find good agreement in all cases,  
except for $z<1$ where the basic assumption of our model is likely to  
break down.

\end{abstract}   
   
\maketitle   
    
\section{INTRODUCTION}      
\label{intro}  
  
In the last decade, a considerable progress has been achieved  
towards the understanding of the nature and evolution of Active  
Galactic Nuclei (AGN). It is now commonly accepted that they are  
powered by Massive Black Holes (MBHs) fuelled by material originating  
in the hosting dark-matter haloes and infalling from the close  
galactic environment in the process described as accretion. New models  
connecting the AGN phenomenon to the hierarchical growth of haloes  
have been developed, based on increasingly robust theoretical  
background substituting phenomenology. In this frame, the masses of  
MBHs are usually assumed to scale with the masses of the hosting  
dark-matter haloes with density estimated from the Press \& Schechter  
(1974) formalism (\eg \citealt{efstathiou88,haiman98}). The AGN  
luminosity is supposed close to the Eddington value during a ``duty  
cycle'' that can be as short as $10^6$ to some $10^7$ years (\eg  
\citealt{haiman98,martini01, wyithe02}) or longer, of the order of  
some $10^8$ years (\eg \citealt{haehnelt93,haehnelt98}). With time,  
these models have become more and more complex and complete. The  
accretion rate onto the MBH varies depending on the model: it can be  
constant or redshift-dependent (\citealt{kauffmann00,haiman00}). The  
AGN activity in many cases appears to be a recurrent event  
(\citealt{siem97,hse01}). The diverging opinions as to this recurrence  
have converged, accounting for tidal interactions and merging  
effects. In addition, the Eddington-limited capability of MBHs to  
accrete mass and the progressive consumption of the hot gas available  
in haloes are also expected to play an important role in the evolution  
of AGN luminosities \citep{cavitt00}.  
  
In several recent papers (\citealt{silkrees98,monaco00,granato01,  
archibald02,volonteri03} and references therein), the joint modelling  
of star formation and the growth of MBHs lying in galactic nuclei has  
been considered in order to account for the observed correlation  
between the masses of the MBH and the hosting spheroid  
\citep{wandel99,woo02}. Spheroids grow through galaxy mergers,  
giving rise to new ellipticals, and through the mass transfer from  
unstable discs (possibly not even settled down) to bulges. Although  
none of these processes is directly associated with the Major Merger  
(MM) of dark-matter haloes, some concatenation of events is believed  
to correlate both phenomena. Firstly, MMs mark the beginning of the  
orbital decay, through dynamical friction, of the more massive  
galaxies of the progenitors, which eventually merge at the center of  
the new halo. Secondly, the shock-heated gas is redistributed and  
begins to cool from the central higher density region at the strongest  
rate ever reached during the whole halo life. Large amounts of gas  
then fall into the central galaxy making its disc easily become  
unstable (particularly at high redshifts where discs are more compact)  
and transfer mass into the bulge. As the spheroid grows some mass may  
reach the nuclear region and feed the central MBH. We therefore  
expect the growth of spheroids and MBHs not only go in parallel, but  
also be punctuated by the rate at which dark-matter haloes form as a  
consequence of MMs \citep{kauffmann00,haiman00}.  
  
It is difficult to tell at which extent this scenario is able to  
reproduce the observed properties of AGN and their dependence with  
redshift. The models found in the literature only address partial  
aspects of it and often include poorly justified assumptions and rough  
estimates of the MBH host density. Nowadays, the development of a  
complete model of this kind has become feasible. In particular,  
accurate analytical expressions for the halo MM rate are available  
(\eg \citealt{percival00,rgs01}) that can be easily implemented. The  
aim of the present paper is to build such a complete, physically  
motivated model and check the validity of the previous scenario by  
comparing the predicted luminosity function of quasars (QLF) with  
the most recent publicly available observations.  
  
Our model is based on the three following ingredients: 1) an accurate  
expression, in terms of halo mass and cosmic time, for the halo MM  
rate (Sec.\ \ref{MMrate}) derived in the frame of the extended  
Press-Schechter model; 2) the evolving properties of MBHs (Sec.\  
\ref{evoMBH}), obtained by solving analytically the governing set of  
differential equations established under a few simple assumptions  
consistent with the parallel growth of MBHs and their hosting galaxies  
from the hot gas available in the halo; and 3) the typical lightcurves  
of AGN associated with MBHs of different masses at different cosmic  
times, inferred under the assumption of an Eddington-regulated  
accretion rate (Sec.\ \ref{QLC}). By comparing the resulting  
theoretical QLFs (Sec.\ \ref{QLF}) with the ones observed at different  
redshifts, the best values of the free parameters entering the model  
are inferred. This leads to a fully consistent analytical expression  
of the QLF (Sec.\ \ref{data}) that recovers both the observed QLF  
shape and its redshift evolution at intermediate and high  
redshifts. In the last section, we discuss the possible shortcomings  
of the model and the implications of our results on the process of  
galaxy formation and evolution. Throughout the present work, we assume  
a $\Lambda$CDM cosmology, defined by the values of the Hubble constant  
(in km s$^{-1}$ Mpc$^{-1}$), density parameter, cosmological constant,  
baryonic fraction, and normalisation of the power spectrum  
respectively given by $(H_0,\Omega\m,\Omega_\Lambda,\Omega\b  
h^2,\sigma _8) = (70,0.3,0.7,0.02,0.75)$, with $h=H_0/100$.

\section{HALO MAJOR MERGER RATE}\label{MMrate}          
         
The Modified Press-Schechter (MPS) model developed by \citet{ssm} and  
\citet*{rgs98,rgs01} is a variant of the extended PS model  
(\citealt{b}; \citealt{BCEK}; \citealt{LC93}) that provides an  
unambiguous definition of the halo formation and destruction times and  
other quantities, such as the halo MM rate, not available in the  
original extended PS model.  
  
By definition, haloes are destroyed in mergers yielding a fractional  
mass increase above some fixed threshold $\delm$; otherwise they  
survive. The formation of new haloes is then consistently defined by  
those mergers in which all partners are destroyed and the whole system  
rearranges. These are the merger events we refer here to as MMs. As shown  
by Raig et al.~(2001), MMs are essentially binary and involve  
similarly massive haloes. More specifically, the halo internal  
structure found in high-resolution \nbody simulations  
\citep{navaetal97} is recovered \citep{metal03} provided $\delm\sim  
0.5$. This is the value of $\delm$ we use hereafter.  
  
In this model, the comoving number density of haloes with masses in the   
range $[M\H$,$M\H+\der M\H]$ at a cosmic time $t\cos$ is given by the   
usual PS mass function   
\begin{eqnarray}          
\nnb         
f(M\H,t\cos)\,\der M\H =   
\qquad\qquad\qquad\qquad\qquad\quad\qquad\qquad\quad\,\,\\         
\sqrt{\frac{2}{\pi}}\frac{\rho_0}{M\H^2}         
\frac{\delc(t\cos)}{\sigma(M\H)}\left |\frac{\der\ln\sigma(M\H)}    
{\der\ln M\H}\right |          
\exp\left [-\frac{\delc^2(t\cos)}{2\sigma^2(M\H)}\right ]\,\der M\H\,,  
\label{umf}          
\end{eqnarray}          
where $\delc(t\cos)$ is the linear extrapolation to the present time  
$t_0$ of the critical over-density for collapse at $t\cos$,  
$\sigma(M\H)$ denotes the r.m.s. mass fluctuation of the density field  
at $t_0$ smoothed over spheres of mass $M\H$, and $\rho_0$ is the  
current mean density of the universe. On the other hand, the merger  
rate at $t\cos$ of progenitors with mass $M\p$ giving rise to  
haloes with masses in the range $[M\H,M\H+\der M\H]$ is \citep{LC93}  
\begin{eqnarray}         
R(M\p\rightarrow M\H,t\cos)\,\der M\H=  
\qquad\qquad\qquad\qquad\qquad\qquad\quad\nnb\\           
\nnb \sqrt{\frac{2}{\pi}}\frac{1}{M\H}\,\frac{\der}{\der t\cos}          
\left[\frac{\delc(t\cos)}{\sigma(M\H)}\right]         
\frac{\der\ln\sigma(M\H)}{\der\ln M\H}         
\left[1-\frac{\sigma^2(M\H)}{\sigma^2(M\p)}\right]^{-3/2}\\         
\times\exp\left\{-\frac{\delc^2(t\cos)}{2\sigma^2(M\H)}\left         
[1-\frac{\sigma^2(M\H)}{\sigma^2(M\p)}\right]\right\}\,\der    
M\H\,.\label{epsmr}         
\end{eqnarray}         
Note that the previous expression corresponds to a transition rate,  
\ie it is defined per halo of mass $M\H$. To obtain the increase per  
unit comoving volume of the total number of haloes with masses  
$[M\H,M\H+\der M\H]$ owing to mergers among progenitors with masses  
$[M\p,M\p+\der M\p]$ one must take $R(M\p\rightarrow M\H,t\cos)\,\der  
M\H$ times $f(M\p,t\cos)\,\der M\p$.  
  
\begin{figure*}  
\centerline{  
\psfig{file=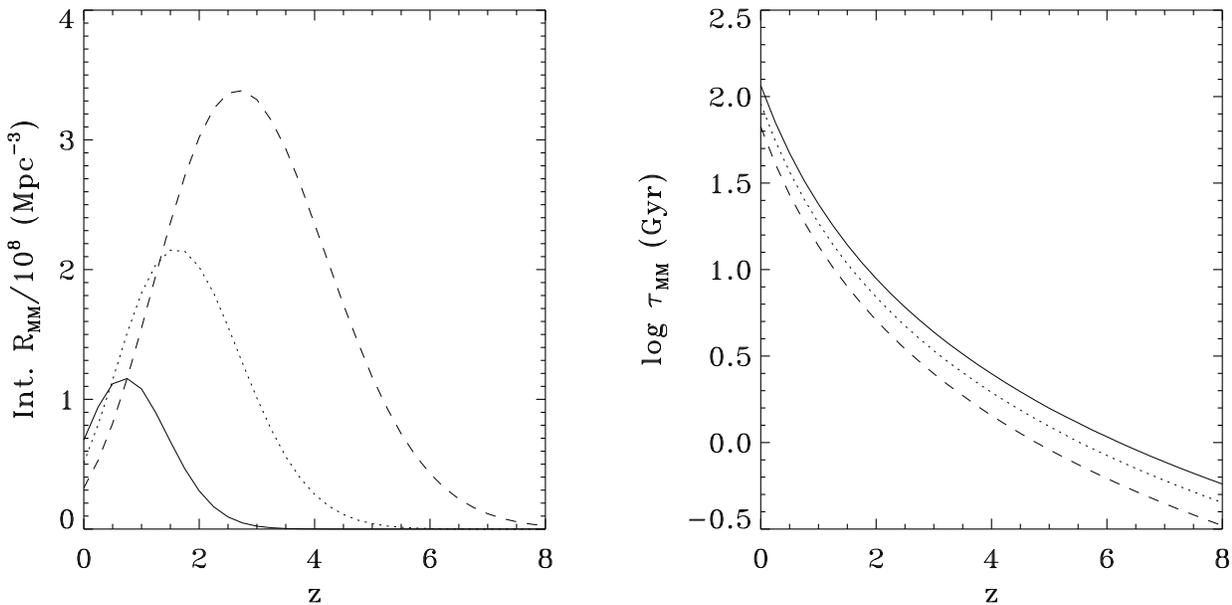,width=18cm,height=9cm}\qquad}  
\caption{MM rate integrated over one 0.5-dex interval in halo mass  
(left panel) and typical interval between two MMs (right panel)  
vs.\ redshift, for three halo masses: $3\times 10^{11}$ (dashed line),  
$3\times 10^{12}$ (dotted line) and $3\times 10^{13}$ M$_{\odot}$  
(solid line). For the integrated MM rate, these masses  
correspond to the lower bounds of the respective integration  
intervals.}  
\label{mmrate}  
\end{figure*}  
  
The required total MM rate per unit comoving volume giving rise to new  
haloes with masses $[M\H,M\H+\der M\H]$ is therefore the integral of  
the latter quantity over the mass of the {\it primary} progenitor in  
the appropriate interval  
\begin{eqnarray}         
R\MM(M\H,t\cos)\,\der M\H =  
\qquad\qquad\qquad\qquad\qquad\qquad\qquad\nnb\\  
\der M\H \int^{M\H/(1+\delm)}_{M\H/2}\,R(M\p\rightarrow M\H,t\cos)\,  
f(M\p,t\cos)\,\der  
M\p\,.         
\label{mmr}         
\end{eqnarray}         
The behaviour of such a MM rate is shown in the left panel of \fig  
\ref{mmrate}. The associated formation rate, $R_{\rm form}$, defined  
again as a transition rate, is then equal to the MM rate given by \eq  
(\ref{mmr}) divided by $f(M\H,t\cos)\,\der M\H$. The goodness of the  
preceding expressions has been checked against \nbody simulations by  
Raig et al.~(2001).  
     
We want to outline the difference between the AGN host  
abundance resulting from the halo MM rate derived here, \eq  
(\ref{mmr}), and that considered in the vast majority of preceding  
models, simply given by the PS mass function, \eq (\ref{umf}). The PS  
mass function implies a positive evolution (\ie it increases with  
increasing redshift) of the comoving number density of haloes with  
masses below the typical mass for collapse. However, this positive  
evolution is not enough for the predicted QLF to reproduce the strong  
positive evolution of the observed QLF. As shown in Appendix  
\ref{kinetic}, the halo MM rate essentially behaves as the product of  
two PS mass functions. This boosts the positive evolution of the  
predicted QLF, which is decisive in making it match the observations.

\section{EVOLVING PROPERTIES OF MBHs}         
\label{evoMBH}     
  
The AGN luminosity depends on the mass $M\BH$ of the associated MBHs  
and the amount of mass $\Delta M\BH$ they accrete (see Sec.\  
\ref{QLC}). In our model, the secular evolution of the mass of a MBH  
corresponds to a sequence of short episodes of intense accretion, with  
characteristic timescale $\tau\acc$ each, between consecutive MMs  
typically separated by $\tau\MM= 1/R_{\rm form}$. Since $\tau\acc$,  
essentially the duty cycle of AGN, is expected to be of the order of  
$10^6-10^9$ yr, we have $\tau\acc\ll\tau\MM$ (for a redshift $z$  
smaller than $\sim 6$, see the right panel of \fig  
\ref{mmrate}). Except for the $\tau\acc$ interval after each MM, the  
accretion rate is thus negligible and the AGN is dormant.  
  
Given the random character of MMs, haloes with the same initial mass  
will evolve in different ways. Similarly, MBHs located in haloes of the  
same mass $M\H$ will be found at different growth phases. Here we are  
not interested in modelling the stair-like (Lagrangian) growth of any individual  
halo or any individual MBH, but the smoother (Eulerian) evolution of the {\it  
average\/} mass at $t$, $M\H(t)$, of haloes with the same initial mass  
and of the {\it average\/} properties at $t$ of MBHs, namely  
$M\BH(M\H,t\cos)$ and $\Delta M\BH(M\H,t\cos)$ in haloes of mass $M\H$.  
  
According to the definition of $\tau\MM$, the Lagrangian evolution of  
the masses $M\H$ and $M\BH$ of objects surviving at $t$ approximately  
satisfy   
\begin{eqnarray}  
\frac{\der M\H}{\der t} \approx  
\frac{M\H(t\cos)}{\tau\MM[M\H(t\cos),t\cos]}\qquad\label{iv} \\  
\frac{\der M\BH}{\der t} \approx {\Delta  
M\BH[M\H(t\cos),t\cos] \over \tau\MM[M\H(t\cos),t\cos]}\,.\label{v}  
\end{eqnarray}  
As shown in Appendix \ref{team}, one can solve analytically equation  
(\ref{iv}) to obtain the evolution of $M\H$ under the effects of MMs  
(we are neglecting the growth of haloes through accretion). To solve  
equation (\ref{v}), however, we need to relate $\Delta M\BH$ with  
$M\BH$. According to \citet{cavitt00}, we must distinguish between the  
Eddington-limited or the supply-limited accretion regimes, depending  
on whether the amount of material afforded by the MBH radiating close  
to the Eddington limit (see Sec.\ \ref{accrmodel}) is larger or  
smaller than the one that can be accreted, respectively.  
  
\begin{figure*}  
\centerline{  
\psfig{file=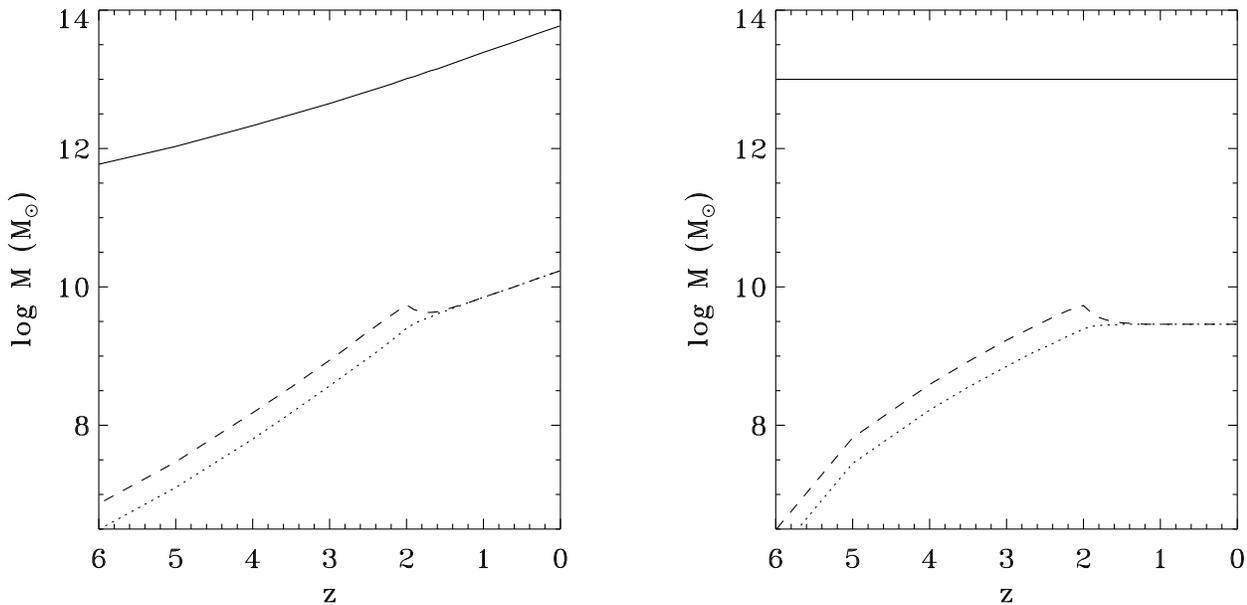,width=18cm,height=9cm}\qquad}  
\caption{Lagrangian (left panel) and Eulerian (right panel) evolution  
of $M\BH$, the instantaneous average mass (dotted line), and $\Delta  
M\BH$, the average mass increase per MM (dashed line), of MBHs located  
within dark-matter haloes with evolving or fixed mass $M\H$ (solid  
line) equal to $10^{13}$ M$_\odot$ at $z\refe=2$ where the change from  
the Eddigton-limited regime to the supply-limited regime takes  
place. The values of the remaining parameters $k$, $\epsilon\gal$ and  
$\epsilon\g$ are those quoted in the third column of Table  
\ref{tabres}.}  
\label{mevol}  
\end{figure*}  
  
As the Eddington luminosity is proportional to the mass of the MBH  
(see eq. [\ref{Edd}]) and the duty cycle can be assumed, in a first  
approximation, with a fixed typical value, in the Eddington-limited  
accretion regime, $\Delta M\BH(M\H,t\cos)$ should be essentially  
proportional to $M\BH(M\H,t\cos)$,  
\begin{eqnarray}  
\Delta M\BH(M\H,t\cos)=k\,M\BH(M\H,t\cos)\label{k}\,.  
\end{eqnarray}  
As shown in Appendix \ref{team}, for a fixed value of $k$ taken as a  
free parameter of the model and some initial conditions set by the  
matching solution in the supply-limited regime, the assumption  
(\ref{k}) leads to analytical expressions for the wanted dependences  
on $M\H$ and $t\cos$ of $M\BH$ and $\Delta M\BH$.  
  
The solution corresponding to the Eddington-limited accretion regime  
is expected to hold at high redshifts where the amount of hot gas in  
haloes is large and its radiative cooling very effective owing to the  
high inner density of haloes. At small enough redshifts, however, a  
substantial fraction of the hot gas in the initial haloes will be  
already consumed and the accretion of MBHs will become  
supply-limited. We then expect the mass accreted by MBHs after the MM  
of their hosting haloes to depend on the small amount of mass accreted  
by the galaxy at the same occasion. The evolution of the two  
quantities $M\BH(M\H,t\cos)$ and $\Delta M\BH(M\H,t\cos)$ is then  
inseparable from that of the masses of the hot gas available in the  
halo, $M\g(M\H,t\cos)$, and the central galaxy, $M\gal(M\H,t\cos)$.  
  
What do we know about the relation between $M\BH$ and $M\gal$? As  
suggested by \citet{kormendy95}, the masses of MBHs and spheroids are  
closely related. \citet{mag98} found a simple proportionality between  
the MBH mass and the luminosity of elliptical galaxies, while  
subsequent works have shown that the relation $M\BH \propto  
\sigma\v^{4.5 \pm 0.5}$, with $\sigma\v$ the bulge velocity dispersion  
of the host galaxy, gives a somewhat tighter correlation  
\citep{fermer00,gebhardt00,merfer01,tremaine02}. Combining the latter  
with the relation observed between $\sigma\v$ and circular velocity  
\citep{ferrarese02}, likely more closely related to the halo  
virial mass, results in a proportionality between the MBH and halo  
masses, $M\BH$ and $M\H$, which is also supported theoretically (see  
\citealt{ferrarese02} and references therein). This latter  
proportionality might be used to infer the evolution with cosmic time  
of $M\BH$, during the supply-limited accretion regime, from that of  
$M\H$ valid at both the Eddington-limited and supply-limited accretion  
regimes.  All the previous observations refer, however, to nearby  
galaxies and any extrapolation to higher redshifts would be very  
risky. On the other hand, a definite time-evolution of both $M\BH/M\H$  
and $M\gal/M\H$ automatically follows, in the supply-limited accretion  
regime, from the assumed parallel growth of $M\BH$ and $M\gal$  
expected to occur as a consequence of halo MMs.  
  
This should take place according to the following simple assumptions  
or approximations. i) The baryonic component of a halo consists  
essentially of hot gas and one main central galaxy. Note that, if the  
two central galaxies in the progenitor haloes merge, the new central  
galaxy will automatically include their baryonic masses. In contrast,  
we neglect those baryons locked in surviving low mass satellites. ii)  
The average mass accreted by the central galaxy after a MM is  
roughly a constant fraction of the hot gas mass $M\g$ available at the  
moment of the MM. iii) The average mass $\Delta M\BH$ accreted by the  
MBH after some delay that can be also neglected in front of  
$\tau\MM$ is a constant fraction of the mass accreted by the hosting  
galaxy as a consequence of the MM or, equivalently, due to assumption  
(ii), a constant fraction $\epsilon\g$ of $M\g$ at the time of  
the MM triggering the whole process. Besides, assumptions (ii) and  
(iii) lead to a constant ratio $\epsilon\gal$ between the average mass  
$M\BH$ of the MBH and the average mass $M\gal$ of the central galaxy,  
as roughly observed. We can therefore write  
\begin{eqnarray}  
M\g(M\H,t\cos)+M\gal(M\H,t\cos)=M\bar(M\H)\label{i}\\  
M\BH(M\H,t\cos)=\epsilon\gal \; M\gal(M\H,t\cos)\qquad\quad\label{ii}\\  
\Delta M\BH(M\H,t\cos)=\epsilon\g\;M\g(M\H,t\cos)\,,\qquad\,\label{iii}  
\end{eqnarray}  
where $M\bar$ is the average baryonic mass in haloes of mass $M\H$,  
equal to the constant ratio $f\bar\equiv\Omega\bar/\Omega\m$ times  
$M\H$, with $\Omega\bar$ and $\Omega\m$ the baryonic and total density  
parameters, respectively, and $\epsilon\gal$ and $\epsilon\g$   
are two free parameters of the model.  
  
The dependences on $M\H$ and $t\cos$ of all previous average  
quantities, in particular of $M\BH$ and $\Delta M\BH$, in the  
supply-limited accretion regime are derived in Appendix \ref{team} by  
solving analytically the set of differential equations implied by the  
assumptions (\ref{i})--(\ref{iii}) and equations (\ref{iv}) and  
(\ref{v}). As an initial condition it is assumed that, at some  
redshift $z\refe$ marking the change between the Eddington-limited and  
the supply-limited accretion regimes, all haloes have essentially the  
same baryonic fraction in the form of hot gas, $\gamma\equiv  
M\g/M\bar=\gamma\refe$. The redshift $z\refe$ is taken as another free  
parameter of the model. In contrast, the match of $M\BH$ and $\Delta  
M\BH$ in the two regimes at $z\refe$ leads to a one-to-one  
correspondence between $\gamma\refe$ and $k$ and there is no need to  
introduce any extra freedom in the model (see Appendix \ref{team}).  
  
The predicted Lagrangian evolution of $M\H$, $M\BH$ and $\Delta M\BH$  
is shown in the left panel of \fig \ref{mevol}. In the plot, we  
represent the results obtained for haloes of $10^{13}$ M$_\odot$ at  
$z\refe$, although all halo masses lead to very similar results. As  
can be seen, $\Delta M\BH$ and $M\BH$, which initially differ by the  
constant factor $k$, become equal shortly after $z\refe$. In the  
latter supply-limited accretion regime, $M\BH$ is also equal to  
$\epsilon\gal\,M\gal$ (by assumption [\ref{ii}]) and approximates to  
$\epsilon\gal\,M\bar$ as $t$ tends to infinity (where all baryons tend  
to be locked in the central galaxy). On the other hand, the trend with  
time of the ratios $M\BH/M\H$ and $\Delta M\BH/M\H$ changes from  
strongly increasing to slightly decreasing in the passage from the  
Eddington-limited to the supply-limited accretion regimes as a  
consequence of the coupling, in the latter, of $\Delta M\BH$ with the  
decreasing gas mass fraction. But what determines the evolution of the  
AGN luminosity is rather the Eulerian evolution of $M\BH$ and $\Delta  
M\BH$ (see Sec. \ref{QLC}), which is represented in the right panel of  
\fig \ref{mevol}. This plot shows that both quantities increase with  
increasing time until $z\refe$, where the values of $M\BH$ begin to  
slightly decrease converging swiftly to those of $\Delta M\BH$,  
forming a plateau for $z$ below $\sim 1.4$.

\section{AGN TYPICAL LIGHTCURVES}         
\label{QLC}         
    
\subsection{Accretion episode}    
\label{accrmodel}      
       
Let us now focus on the typical evolution, {\it during one individual   
accretion episode\/}, of a MBH with initial mass, at time $t\cos$,   
equal to the average mass of MBHs in haloes of mass $M\H$ at that  
moment, accreting a total mass of gas also equal to the average   
value. In order to clearly distinguish between the initial mass of the   
MBH, given by $M\BH(M\H,t\cos)$ (varying in a timescale $\tau\MM$),   
and the mass of the MBH during the accretion episode (varying in a   
characteristic timescale $\tau\acc \ll \tau\MM$), we denote this latter   
simply by $M$, without subindex BH.   
   
As mentioned, some time after a MM of haloes we expect radial infall  
of material to start feeding the MBH at the centre of the new  
halo. This central MBH can be the result of the merger of the two MBHs  
in the respective progenitor haloes, since the respective central  
galaxies can merge and their central MBHs may eventually fall into the  
centre of the new spheroid (also due to dynamical friction within the  
new galaxy) and merge. But only one of these two MBHs may finally be  
found at the nucleus of the central galaxy. In any event, we do not  
need to consider the frequency of MBH mergers since the initial mass  
assumed for the accreting MBH, equal to the average mass of MBHs in  
haloes with $M\H$ at $t\cos$, implicitly accounts for it.  
   
\begin{figure*}  
\centerline{  
\psfig{file=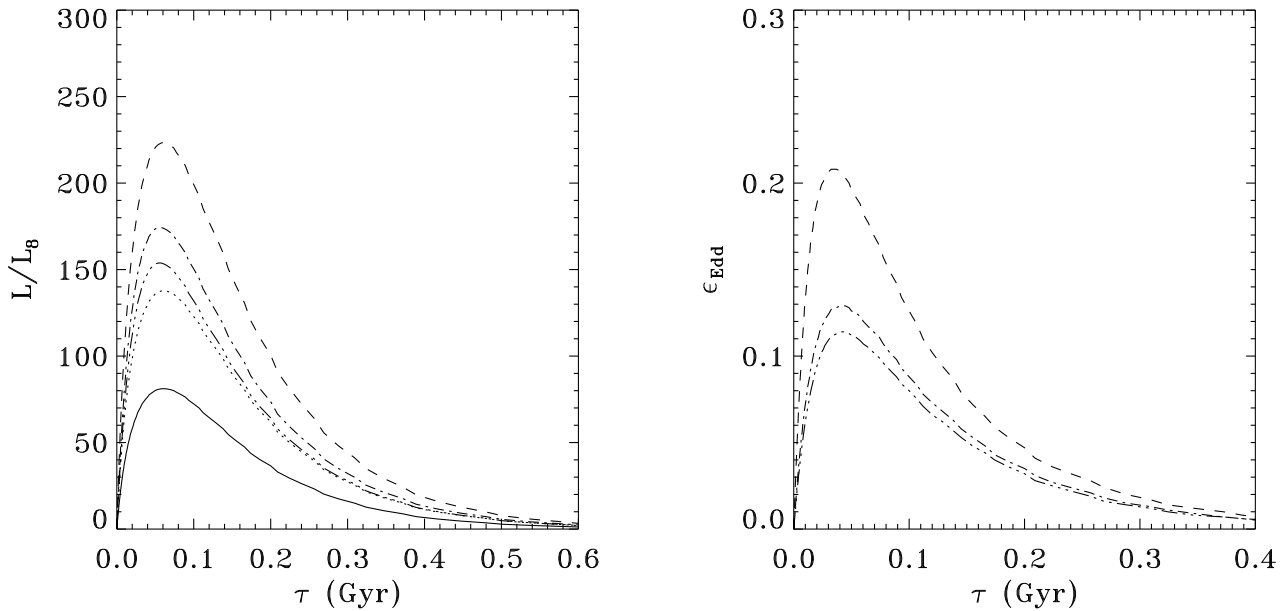,width=18cm,height=9cm}\qquad}  
\caption{Predicted AGN lightcurves (left panel) and corresponding  
Eddington efficiencies (right panel) during an accretion episode onto  
the average MBH within haloes of typical mass $10^{13}$ M$_{\odot}$ at  
redshifts equal to 3.0 (full line), 2.6 (dotted line), 2.2 (dashed  
line), 1.7 (dot-dashed line) and 0.3 (triple dot-dashed line). In the  
case of $\epsilon\Edd$ we only plot the solutions for $z\le z\refe$ since  
the remaining curves overlap with that of $z\refe=2$. The values of  
all the parameters used are given in the third column of Table  
\ref{tabres}.}  
\label{Lt}         
\end{figure*}      
            
Material is expected to be supplied to the central MBH, initially in  
an increasing phase, then in a decreasing one, during a time of the  
order of $\tau\acc$. Therefore, we assume that the accretion rate  
$\dot M$ follows a bell-shaped form  
\begin{equation}  
\dot M(\tau) =\Delta M\BH(M\H,t\cos)\;     
\frac{m'(\theta)}{\tau\acc}\label{e1}  
\end{equation}  
determined by the dimensionless phenomenological function  
\begin{equation}  
m'(\theta)\equiv\frac{4}{3}\exp\left(-\theta\right)          
\left[1-\exp\left(-3\theta \right)\right]\,,\label{acc1}         
\end{equation}  
with $\tau$ the time elapsed since the beginning of accretion $t\cos$,  
$\theta\equiv \tau/\tau\acc$, and a dot denoting differentiation over  
$\tau$. The {\it typical\/} evolution of the mass $M$ of the MBH  
during such an accretion episode then follows by integration of \eqs  
(\ref{e1}) and (\ref{acc1})  
\begin{equation}  
M(\tau)=M\BH(M\H,t\cos) + \Delta M\BH(M\H,t\cos) \; m(\theta)\,,         
\label{e2}  
\end{equation}  
with $m(\theta)$ equal to the integral of $m'$ from 0 to  
$\theta$  
\begin{equation}        
m(\theta)\equiv 1-\frac{4}{3} \; \exp(-\theta)\;+\;\frac{1}{3}   
\exp(-4\theta)\,.\label{mdet}   
\end{equation}   
  
$\Delta M\BH$ and $M\BH$ in \eqs (\ref{e1}) and (\ref{e2}) are the  
evolving average properties of MBHs introduced in Section  
\ref{evoMBH}. The value of $\tau\acc$, which should be related to the  
dynamical properties of the MBH close environment, is difficult to  
tell. For this reason, we take $\tau\acc$ as another free parameter of  
the model.  
  
\subsection{Radiation model}    
\label{radmodel}      
      
The accretion of gas onto the MBH reactivates the AGN. An accurate  
model of the resulting lightcurve would require a relativistic  
approach. But this is outside the scope of the present study,  
aimed merely at understanding the main trends of the observed QLF and  
its redshift evolution. For this reason we shall neglect any  
relativistic correction.  
    
Under the classical black hole paradigm, a source with mass $M$    
powered by spherical accretion can reach a maximum luminosity, called    
Eddington luminosity, given by   
\begin{eqnarray}         
L\Edd=\frac{4\pi\,c\,G\,\rho\acc M}{\sigma \Th C\Edd \,n_{\rm e}}\,,   
\label{Edd}  
\end{eqnarray}         
with $G$ the gravitational constant, $c$ the speed of light, $\rho\acc$    
and $n_{\rm e}$ the mass density and electron density, respectively,    
of the accreted matter, $\sigma\Th$ the Thompson cross-section and    
$C\Edd\simeq 0.2$ a dimensionless constant which accounts for the    
variations with electron energy of the Compton cross-section, averaged    
with a standard quasar SED \citep{ciotti01}.    
       
If, at some radius $r$ at a given time $\tau $, the luminosity exceeds  
the Eddington limit, the radiative acceleration will exceed the  
gravitational acceleration and accretion will stop. This will cause  
the radiation to stop soon after, allowing accretion to start again,  
and so on. In this way, accretion and radiation will oscillate on a  
characteristic timescale of the order of the infall time at $r$. In  
fact, provided that this characteristic time is short enough, the  
luminosity of the AGN will tend to adjust to the accretion rate set  
by external conditions linked to the MM event, thus reaching a  
(possibly time-averaged) quasi-stationary regime. The characteristic  
timescales of light variations observed in AGN (years or less) are  
considerably shorter than the timescale of accretion onto the MBH (at  
least $10^{6}$ yr). Hence, such a quasi-stationary regime should be  
reached swiftly during accretion.  
  
In these circumstances, the radiative pressure at any radius $r$  
results in a reduced effective gravity  
\begin{eqnarray}      
g\eff (r)=-\frac{GM}{r^2}\left(1-\epsilon\Edd\right)\,,  
\label{mvmnt}   
\end{eqnarray}    
which we have expressed in terms of the Eddington efficiency  
$\epsilon\Edd\equiv L/L\Edd$. The bulk of the energy release comes  
from the region with maximum accretion efficiency, which occurs at a  
radius   
\begin{eqnarray}   
r\last=k\last\frac{2\,GM}{c^2}\,, \label{rlast}   
\end{eqnarray}   
corresponding to the last marginally stable Keplerian orbit  
\citep{chakrabarti00}, with $2\,GM/c^2$ the Schwarzschild radius and  
$k\last$ a proportionality factor typically ranging between one half  
and two or three, depending on the form of the orbit and the angular  
momentum (hence, the metric, either Schwarzschild or Kerr) and the  
topology (\eg simple, binary, toroidal) of the MBH. In the present  
calculations, we do not distinguish however among all these possible  
configurations and adopt the intermediate value of $k\last  
=1.5$. According to \eq (\ref{mvmnt}), the terminal infall velocity  
$v\last$ for accreted material is, therefore,  
\begin{eqnarray}      
v\last^2 =2\,\frac{G\,M}{r\last} (1-\epsilon\Edd)\,.  
\label{vterm}       
\end{eqnarray}  
Since the bolometric luminosity is the kinetic energy of the accreted  
material converted into radiation, we have  
\begin{eqnarray}      
L = \frac{1}{2}\, \dot M \, v\last^2\,,         
\label{Lbol}       
\end{eqnarray}      
and, by substituting \eqs (\ref{rlast}) and (\ref{vterm}) into \eq  
(\ref{Lbol}) and dividing it by $L\Edd$, we find  
\begin{eqnarray}   
\epsilon\Edd=\left[1+ \frac{8\pi\,k\last\, G \rho\,\acc M}  
{c\,\sigma\Th\,C\Edd\,n_{\rm e}\,\dot M} \right]^{-1}\,.\label{edd}   
\end{eqnarray}    
Finally, by taking into account the definition of the Eddington   
efficiency $\epsilon\Edd$ given above, we are led to the following  
non-linear expression for the AGN bolometric luminosity  
\begin{eqnarray}     
\frac{L_8}{L} =\frac{M_8}{M} + \frac{\dot M_8}{\dot M}\qquad\qquad \qquad    
\label{ldet}      
\end{eqnarray}    
in terms of the two functions, $M$ and $\dot M$, defining the accretion   
episode (eqs. [\ref{e1}] -- [\ref{mdet}]) and the three constants  
\begin{eqnarray}    
M_8\equiv 10^8\, {\rm M_{\odot}}\nnb   
\qquad\qquad\qquad\qquad\qquad\qquad\\   
L_8\equiv\frac{4\pi c\, G  
\rho\acc M_8}{\sigma\Th C\Edd n_{\rm e}}=6.8\times 10^{46} \, {\rm  
ergs/sec}\nnb\; \\   
\dot M_8 \equiv 2k\last\frac{L_8}{c^2} \,=3.6\   
{\rm M}_{\odot}/{\rm yr}\,.\quad\qquad\qquad\nnb  
\end{eqnarray}  
  
The resulting typical AGN lightcurves, $L(\tau)$, and the  
corresponding Eddington efficiencies, $\epsilon\Edd(\tau)=L(\tau)/L_8  
\times M_8/M(\tau)$, are shown in \fig \ref{Lt} for the same fixed  
halo mass plotted in \fig \ref{mevol} and several illustrative  
redshifts. Note that, in our model, $\epsilon\Edd$ is not constant  
along the duty cycle but, like the bolometric luminosity, it is a  
bell-shaped function of time taking values substantially lower than  
unity. The maxima of both $L$ and $\epsilon\Edd$ depend on the  
Eulerian evolution of $M\BH$ and $\Delta M\BH$ (see the right panel of  
\fig \ref{mevol}) through \eqs (\ref{e1}) and (\ref{e2}). From \eq  
(\ref{edd}), we have that $\epsilon\Edd$ is constant as long as  
$\Delta M\BH$ is proportional to $M\BH$, namely, during the whole  
Eddington regime and at redshifts below $\sim 1.4$. Between these two  
phases, right after $z\refe$, it diminishes with increasing time. On  
the other hand, from \eq (\ref{ldet}) we have that, as long as $\Delta  
M\BH$ is proportional to $M\BH$, $L$ is also proportional to  
them. Therefore, the luminosity increases with time during the whole  
Eddington-limited accretion regime and diminishes, after $z\refe$, to  
a constant value where $M\BH$ and $\Delta\BH$ become equal. This  
implies a negative luminosity evolution of AGN associated with haloes  
of a fixed mass during the Eddington-limited accretion regime, a positive  
evolution in the initial phase of the supply-limited accretion regime  
until $z\sim 1.4$ and a null evolution at lower redshifts. It is worth  
noting that the luminosity also increases with increasing halo mass,  
but not the Eddington efficiency.  
    
\section{AGN LUMINOSITY FUNCTION}\label{QLF}          
           
The (differential) AGN luminosity function, $\phi(L,t\cos)$, is  
defined through the relation  
\begin{eqnarray}     
\der^2 N(L,t\cos)= \der n(L,t\cos)\;\der V = \phi(L,t\cos)\;\der     
L\;\der V\,,   
\end{eqnarray}    
where $N(L,t\cos)$ and $n(L,t\cos)$ are the number and comoving number  
density, respectively, of AGN with absolute bolometric luminosity  
$L$ at a given time $t\cos$ and $\der V$ is the element of comoving  
volume. Had all AGN the same typical strictly increasing  
(or decreasing) lightcurve $L(\tau)$, we would have  
\begin{eqnarray}    
\phi(L,t\cos) \equiv \frac{\der n(L,t\cos)}{\der L}=\nnb    
\qquad\qquad\qquad\qquad\qquad\qquad\qquad\;\\    
\pm \lim_{\Delta L\rightarrow 0}\;\frac{1}{\Delta L}   
\left[\int^{t\cos}_{{t\cos}-\tau(L)} \hskip -10pt R\qb(t')\;    
\der t' - \int^{t\cos}_{{t\cos}-\tau(L+\Delta L)}    
\hskip -25pt R\qb(t')\; \der t'\right]=\nnb\,\,\\    
\pm \lim_{\Delta L\rightarrow 0}\;\frac{1}{\Delta L}    
\int^{\tau(L+\Delta L)}_{\tau(L)} R\qb(t')\,    
\der t'\approx \nnb\qquad\qquad\qquad\quad\,\\    
\pm \lim_{\Delta L\rightarrow 0}\, R\qb(t\cos)    
\left[\frac{\tau(L+\Delta L)-\tau(L)}{\Delta L}\right]=    
\pm R\qb(t\cos)\frac{\der \tau}{\der L},\;\;\;    
\label{basic}    
\end{eqnarray}    
where $R\qb$ is the AGN reactivation rate, the + ($-$) sign is for a  
strictly increasing (decreasing) lightcurve and the two integrals on  
the right of the first equality give the comoving number density of  
AGN with luminosity larger (smaller) than $L$ and $L+\Delta L$,  
respectively, at $t$. Note that, in the derivation of \eq  
(\ref{basic}), we have taken into account the fact that the  
characteristic accretion time $\tau\acc$ is much shorter than the  
characteristic time between MMs $\tau\MM$, so that, along the time  
interval of integration, $R\qb$ is approximately constant and equal to  
$R\qb(t\cos)$.   
  
Therefore, in the more realistic case of AGN with typical  
lightcurves dependent on halo mass $M_H$, we can write  
\begin{eqnarray}    
\phi(L,t\cos) =    
\int_0^{\infty}\der M\H\, R\qb(M\H,t\cos)    
\left[\frac{\der \tau_1}{\der L}- \frac{\der \tau_2}{\der L}\right]         
(L,M\H,t\cos)\,,  
\label{phiM}    
\end{eqnarray}  
where $R\qb(M\H,t\cos)\,\der M\H$ is the reactivation rate of AGN  
in haloes with masses in the range $[M\H,M\H+\der M\H]$ and where we  
have accounted for the fact that, as shown in \fig \ref{Lt}, each  
typical quasar lightcurve has two branches, one increasing and the other  
decreasing with increasing $\tau$ (indexes 1 and 2, respectively).  
  
Hence, to infer the QLF predicted by our model at any time or  
redshift, we must simply write, in \eq (\ref{phiM}), the quasar  
reactivation rate in terms of the halo MM rate,  
\begin{equation}  
R\qb(M\H,t\cos)=\nu\,R\MM(M\H,t\cos)\,,  
\label{nu}  
\end{equation}  
with $R\MM$ given in \eq (\ref{mmr}), and take the functions  
$\tau_i(L,M\H,t\cos)$ as the inverse of the increasing and decreasing  
parts of the typical luminosity curve $L(\tau,M\H,t\cos)$ given in \eq  
(\ref{ldet}). The proportionality factor $\nu$ in \eq (\ref{nu})  
accounts for possible inaccuracies in the assumptions adopted as well  
as for possible biases affecting the observational data. For example,  
although our model presumes one single quasar event per halo MM, this  
might deviate from reality (factor ${N_{\rm QSO}}/{N\MM}$ in \eq  
[\ref{Ni}]) as there are indications that haloes may harbour more than  
one quasar (\citealt{mwf99}; see also \citealt{croom01b}). On the other  
hand, there are also reasons to introduce such a normalisation factor  
from the observational point of view. More specifically, no sample is  
ever free of incompleteness effects and selection biases (factors  
$\mathcal{C}$ and ${N_{\rm QSO,obs}}/{N_{\rm QSO,tot}}$ in \eq  
[\ref{Ni}], respectively). Last but not least, one must take into  
account, in addition, the orientation of detected quasars (factor  
${2\Omega\em}/{4\pi}$ in \eq [\ref{Ni}]). All these effects should not  
present a strong evolution in the range $0 < z < 2.5$, but be rather  
constant (with eventual variations in some specific redshift ranges  
due to redshift-dependent observational biases). As several of these  
effects are rather uncertain, we take  
\begin{eqnarray}     
\nu=\mathcal{C} \times \frac{2\Omega\em}{4\pi}   
\times \frac{N_{\rm QSO,obs}}{N_{\rm QSO,tot}}     
\times \frac{N_{\rm QSO}}{N\MM}    
\label{Ni}  
\end{eqnarray}  
as another free parameter of the model.  
         
\section{COMPARISON TO OBSERVATIONS}          
\label{data}          
  
\begin{figure*}  
\centerline{  
\psfig{file=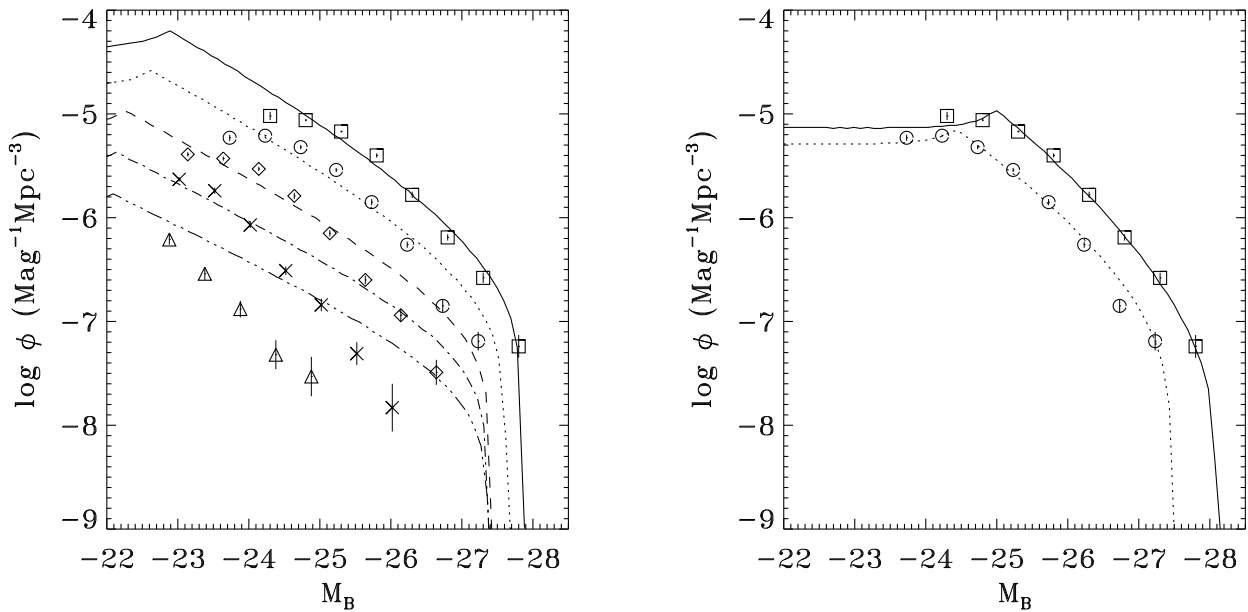,width=18cm,height=9cm}\qquad}  
\caption{Predicted QLFs (lines) compared to the observed blue QLFs  
drawn from the 2QZ sample (symbols) at various redshifts. The  
tiny error bars affecting the observational data are only  
statistical. Left panel: redshifts in the $[0.3,2.3]$ range;  
quasars are binned around $\langle z \rangle$=0.435, 0.730, 1.080,  
1.808, and 2.020 (bottom to top). Right panel: restriction to the two  
bins with $\langle z \rangle$ larger than unity where the model gives  
the best fit. The values of the parameters used are given in Table  
\ref{tabres}, columns 2 and 3, respectively. For clarity, the QLFs are  
shifted upwards by 0.2 units with each increasing $\langle z  
\rangle$.}  
\label{logphi}    
\end{figure*}    
  
As explained in Appendix \ref{team}, we do not expect $z\refe$ to  
substantially vary with halo mass, but the situation is less clear for  
the other quantities taken as constant parameters in the present  
model. For instance, by combining the scaling relation (\ref{sp}) with  
the observed $M\BH - \sigma\v$ correlation, we could infer an  
expression for $\epsilon\gal$ as a function of halo mass and  
redshift. However, letting these quantities to depend on $M\H$ and  
$t\cos$ or on any of these variables alone would greatly complicate  
the task of obtaining an analytical model and would introduce too much  
freedom in the problem. Thus, we prefer to assume them having  
fixed values in a first approximation.  
  
\subsection{Constraints to the parameters}     
  
So far we have implicitly assumed that all haloes, irrespective of  
their mass, harbour MBHs evolving according to the model given in  
Section \ref{evoMBH} and that all MBHs radiate according to the model  
described in Section \ref{QLC}. But this may not be the case. Only  
haloes in some finite mass range are likely to feed MBHs at the suited  
rate and reach the appropriate radiative efficiency (\ie the  
capability of converting the kinetic energy of the matter falling at  
$r\last$ into radiation). We should therefore also examine the effects  
of adopting a limited halo mass range, $[M\min,M\max]$. In summary,  
the free parameters of the model are: $z\refe$, $\epsilon\gal$,   
$\epsilon\g$, $\tau\acc$, $k$, $M\min$, $M\max$ and $\nu$.  
  
To find the best values of these parameters we have followed an  
iterative $\chi^2$-minimisation fitting procedure to the observed QLF,  
with each free parameter of the model taking values from a  
pre-constructed grid within certain initial intervals (see below). For  
the following iterations, however, the width of these intervals is  
progressively reduced and their centres are re-defined using the  
best-fitting parameter values of the former step in such a way that  
the free parameters are allowed to take values outside their initial  
intervals. (Actually, the only parameters fitted in this way  
are $z\refe$, $\epsilon\gal$, $\epsilon\g$, $\tau\acc$, $k$, $M\min$  
and $M\max$; the value of $\nu$ is drawn, at each step, by taking the  
best simultaneous normalisation of the predicted QLFs.)  
  
The redshift $z\refe$ marking the change from the Eddington-limited to  
the supply-limited accretion regimes is initially allowed to take  
values in the range [0,10]. The ratio between the masses of MBHs and  
their central galaxies, $\epsilon\gal$, is allowed to vary in the  
range $[5.5\times 10^{-4},6.7\times 10^{-3}]$ indicated by  
observations \citep{kormendy01,mclure01,wandel02,falomo02,wu02}.  
Contrarily to $\epsilon\gal$, the fraction of gas in the host galaxy  
that is accreted by the MBH at the occasion of a MM, $\epsilon\g$, is  
poorly constrained. So we leave $\epsilon\g$ completely free. The  
timescale $\tau\acc$ is supposed to be in the range [$10^6,10^9]$ yr  
bracketing the values mentioned in the literature for the duty cycle  
of AGN. The previous bounds for $\tau\acc$ together with \eq  
(\ref{edd}) under the approximation $\dot M=\Delta M/(2\tau\acc)=k  
M/(2\tau\acc)$ lead to a fractional mass increase, $k$, of the MBH  
during an Eddington-limited accretion episode in the range  
[0.05,50]. $M\min$ and $M\max$ are assumed to lie initially in the  
ranges $[10^{11},10^{12.5}]$ M$_{\odot}$ and $[10^{13},10^{14.5}]$  
M$_{\odot}$, respectively. Finally, the normalisation factor $\nu$ is  
also left completely free.  
  
\subsection{Results for the 2dF QSO sample}\label{param}  
  
Quasars are not always easy to distinguish from other types of  
objects. Moreover, the different samples gathered in different  
wavelengths are usually small. In this concern, the 2dF quasar  
redshift (2QZ) sample \citep{croom01a} seems the most  
appropriate for our purposes. The 2QZ sample is very large, uniformly  
selected from well-defined criteria and of high overall completeness  
$\mathcal{C} \sim 93\%$ to the limiting magnitude $b_J=20.8$, after  
corrections for incompleteness (see http://www.2dfquasar.org/). The  
subsample of 2QZ used here comprises some 10000 colour-selected,  
moderately faint (M$_{\rm B} < 21)$ quasars, covering the redshift  
range [0.3,2.3]. The candidates were all point-like objects selected  
by their $UV$ excess on a $(U-B) \times (B-R)$ plane, introducing an  
abrupt decrease of the efficiency of the selection both at $z>2.3$  
($UV\hskip -1.5pt X$ criterion) and at $z<0.3$ (due to morphological  
classification). To perform the fit, the corresponding QLFs for  
several redshift bins derived by \citet{croom01a} assuming $H_0=50$ km  
s$^{-1}$ Mpc$^{-1}$, $\Omega_m=1$ and $\Omega_\Lambda=0$ have been  
converted (in luminosities and densities) to the $\Lambda$CDM  
cosmology used here and the bolometric luminosities predicted by our  
model transformed to absolute luminosities in the blue band by  
considering a constant ratio $L_{\rm B}/L_{\rm Bol}=0.023$, inferred  
from a typical quasar SED \citep{elvis94}.  
  
\begin{figure}  
\centerline{  
\psfig{file=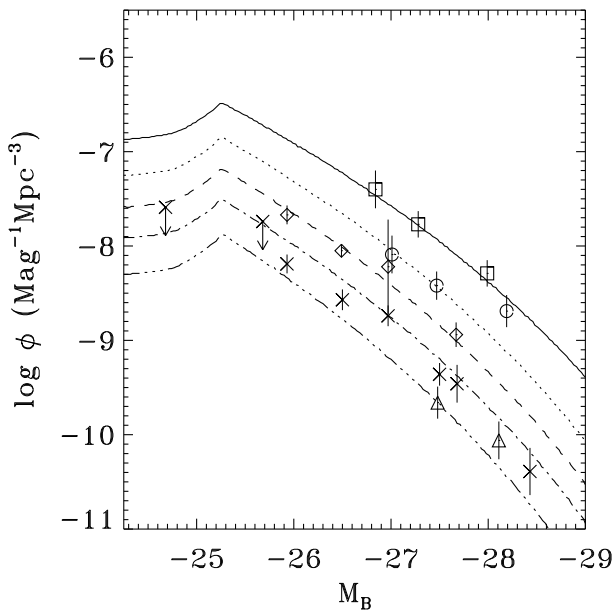,width=9cm,height=9cm}\quad}  
\caption{Predicted and observed QLFs for quasars in the very high-$z$  
range $[3,4.7]$ binned around $\langle z \rangle=3.75$ (solid line and  
squares), 4.15 (dotted line and circles), 4.3 (dashed line and  
diamonds), 4.4 (dot-dashed lines and crosses) and 4.7 (triple  
dot-dashed lines and triangles). The values of the parameters used are  
given in column 4 of Table \ref{tabres}. Both predictions and data are  
shifted downwards by 0.3 units with each increasing $\langle z  
\rangle$. Note that the two first points on the left are only upper  
limits.}  
\label{fan}         
\end{figure}      
  
In the left panel of \fig \ref{logphi}, we plot the predicted and  
observed QLFs for the best-fitting values of the parameters, listed in  
the second column of Table \ref{tabres}. In the third column of the  
Table, we also quote the best values of the parameters for the data  
restricted to $1.27<z<2.3$, where the model gives a much better fit  
(see the right panel of \fig \ref{logphi}). The possible origin of the  
poor fit at low redshifts is discussed in Section \ref{discuss}. As  
can be seen from Table \ref{tabres}, all parameters take values in the  
respective expected ranges. The value of $M\min$ quoted in the  
second column is, however, essentially an upper limit since the lower  
error bound is not constrained by the data. This is because varying  
this lower mass boundary essentially modifies the fainter part of the  
luminosity function while, at all redshifts, the cut-off due to the  
limiting magnitude $b_J=20.8$ is effective at magnitudes brighter than  
the cut-off due to $M\min$. Besides, at low $z$ there is another  
cut-off at $M_{\rm B}=-23.0$ due to the selection of stellar  
morphologies. A more realistic model, likely giving better fits to the  
observed QLFs at the faint end, would require a progressive decline of  
the integrand in equation (\ref{phiM}) beyond some halo mass instead  
of a sharp cut-off at $M\min$, although at the cost of increasing the  
number of free parameters. In this concern, the best value of $M\min$  
should be taken with caution.  
  
The factor ${N_{\rm QSO,obs}/{N_{\rm QSO,tot}}}$ appearing in the  
decomposition of $\nu$ given in \eq (\ref{Ni}) reflects two combined  
effects: the intrinsic properties of the objects and selection  
biases. The 2QZ quasars are blue objects with redshifts lower than  
$\sim$2.3, selected by their $UV$ excess. Therefore, this factor  
represents here the number of low to intermediate blue quasars over  
the total quasar population. Several recent results based on X-ray  
observations (\eg \citealt{rosati02}) show that the reddened, or
{\it type-II}, quasars  
could be as many as the blue ones detected in the optical  
passbands and, therefore, $UV\hskip -1.5pt X$ quasars could be only a  
fraction (but a large one) of the overall optical quasar population,  
estimated to be of the order of 50$-$90\%. 
The factor ${2\Omega\em}/{4\pi}$, accounting for the  
orientation effects, is $\sim0.25$ according to the unified models  
(see \eg \citealt{elvis02}). As mentioned in Section \ref{QLF},  
${N_{\rm QSO}}/{N\MM} \sim 1$, although (slightly) greater values are  
theoretically possible. Accordingly, a most crude estimate of the  
value of $\nu$ gives an upper limit of 10\% to within a factor of  
around 2. Thus, the predicted value of $\sim 10\%$ also lies within  
the expected range.  
  
\subsection{Results for QSOs at very high redshifts}\label{highzparam}  
    
We have also compared the predictions of our model to the very  
high redshift QLF, compiled from the points of \citet{fan01},  
\citet{ken95}, \citet{ken96} and \citet{schmidt95}. The best fit is  
once again satisfactory (see \fig \ref{fan}), although the values of  
some parameters move substantially apart from those previously found  
(compare column 4 with column 3 of Table \ref{tabres}). However,  
taking into account the large uncertainty affecting the new values, we  
see that such differences are not significant. In fact, the results of  
this high-$z$ fit must be taken with caution because of possible  
systematic effects, such as lensing, affecting the data (see \eg  
\citealt{wyithe02}). Moreover, the small error bars associated with  
the values of the parameters drawn from the 2QZ sample are likely  
underestimated since they only account, once again, for statistical  
errors in the observed QLFs, almost negligible in this case owing to  
the large number of quasars included in each bin. The only parameters  
that, despite all, seem to change are $\tau\acc$ and, to a  
lesser extent, $M\min$. As mentioned, we should not attach too much  
importance to the exact value of this latter parameter as a smooth  
cut-off in halo mass is likely more realistic than a sharp  
one. Concerning $\tau\acc$, the tendency it shows to decrease with  
increasing $z$ was foreseeable since all physical sizes and  
separations become smaller and, hence, all dynamical processes go  
faster at higher redshifts. In particular, the typical interval  
between MMs diminishes with increasing $z$ (see \fig \ref{mmrate}),  
thus $\tau\acc$ must also do so if the condition $\tau\acc\ll \tau\MM$  
is to be satisfied.  
  
\begin{figure*}  
\centerline{\psfig{file=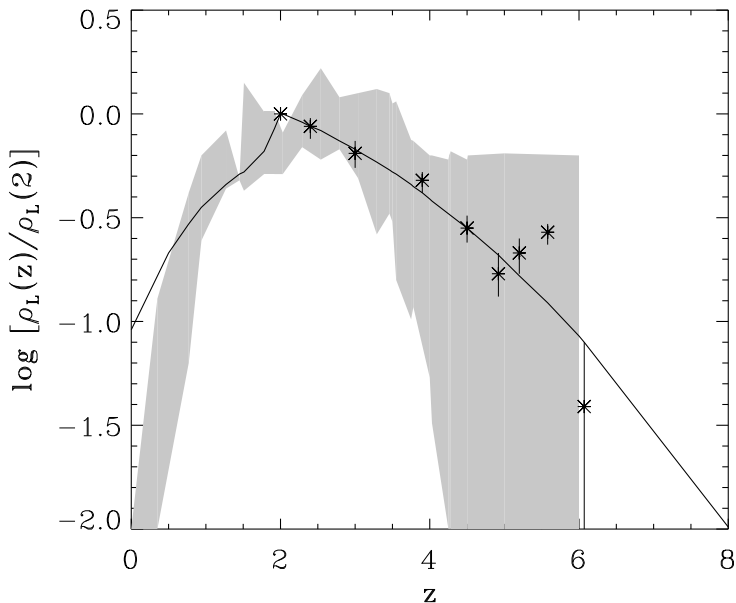,height=9cm}}  
\caption{Integrated quasar light density vs.\ redshift compared to data  
in optical \citep{shaver96} and X-rays \citep{miyaji00}, and the ionising  
flux estimated from the absorption spectra of high-$z$ quasars   
\citep{mcdo01}. The model predictions are computed for the set  
of parameters shown in Table \ref{tabres}, column 5. Data  
and model predictions are normalised to $z=2$.}  
\label{rhoqlz}    
\end{figure*}    
  
\subsection{Results for the integrated quasar light}    
\label{RhoL}  
  
An alternate way of estimating the best values of the parameters is  
matching the quasar light density over the redshift range [0,6]. Note,  
however, that the model is very poorly constrained by the observations  
at $z>4$ owing to the large fraction of the quasar population that is  
likely to be dust-enshrouded \citep{xia01,rosati02,norman02}.  \fig  
\ref{rhoqlz} compares the model predictions with available data (the  
gray-shaded region). The lower limit of the observations arises from  
the optical data by \citet{shaver96}, while its upper limit is deduced  
from X-ray measurements (\eg \citealt{miyaji00}). We also include the  
data points coming from an estimate of the photoionisation rate based  
on the observed absorption spectra in high-$z$ quasars (\fig 3 of  
\citealt{mcdo01}). In fact, the data used to adjust our model are the  
latter (normalised at $z=2$) since they are better adapted to our  
$\chi^2$ fitting procedure. Caution must be paid, however, to the fact  
that the total ionizing flux is believed to be dominated, at  
$z>4$, by that of the blue stellar population.  
  
The predicted behaviour of the quasar light density is compatible with  
the observed one for intermediate and high redhifts, while there is,  
once again, less evolution than observed at $z<1$. Interestingly  
enough, the best values of the parameters obtained from the fit to the  
estimated ionizing background, given in column 5 of Table  
\ref{tabres}, are in general close to and, in any case, compatible  
with those obtained from the 2QZ sample at $z>1$. The only exceptions  
are $\tau\acc$ and $M\min$. The former takes, not surprisingly, an  
intermediate value between those quoted in columns 3 and 4. On  
the other hand, the slightly smaller value of the latter is even more  
consistent than the previous ones with the lower boundary of the observed mass range of current MBH  
masses, $10^6<M\BH/M_{\odot}< 10^{10}$, which corresponds to $10^{9.6}<M\H<10^{13.6}$  
at the limit $t \rightarrow \infty $   
where $M\BH \rightarrow \epsilon\gal M\gal \simeq \epsilon\gal f\b M\H$.
This overall good agreement might indicate  
that quasars are the major responsible of the ionizing background at  
$z<4$ where the fit is very good or, alternatively, that the ratio of  
the contributions to this background of quasars and young stars  
remains approximately constant in that redshift interval.

\section{DISCUSSION}    
\label{discuss}   
  
We have developed a fully consistent analytical model of QLF from  
reasonable physical assumptions and used available data on the QLFs at  
various redshifts to fix the values of the free parameters. The  
predicted QLF agrees with the available observations at redshifts  
higher than $\sim 1$.  
  
The fact that, the more accurate halo MM rate used in our model  
behaves essentially as the product of two PS mass functions instead of  
just a single one as often adopted explains the more marked positive  
evolution shown by the inferred QLF. Yet, the evolution so predicted  
is not strong enough to match the observations. The QLF calculated  
from the adopted MM rate but assuming a MBH mass proportional to the  
halo mass and the associated quasar radiating close to the Eddington  
regime fails to reproduce the observed data. A successful approach  
requires also modelling the typical evolution on cosmological  
timescales of MBHs and the typical lightcurve of AGN. In the present  
study, such a theoretical lightcurve rests on the assumptions of a  
phenomenological bell-shaped accretion rate and a physically grounded  
self-regulated radiation, while the modelling of the MBH growth is  
based on reasonable assumptions emanating from the expected connection  
between halo MMs and the reactivation of AGN.  
  
\begin{table*}  
\caption{Best-fitting parameters, column 1, corresponding to four data  
sets. In columns 2 and 3, the observed QLFs drawn from the 2QZ survey  
in the two redshift intervals quoted. In column 4, the high-$z$ QLFs  
drawn essentially from the SDSS survey \citep{fan01} and, in column 5,  
the integrated quasar light estimated from the absorption spectra of  
high-$z$ QSOs \citep{mcdo01}. $z\refe$ is the redshift at which the  
accretion regime changes from Edington to supply-limited. $k$ is the  
fractional mass increase of the MBH in the Edington-limited  
regime. $\epsilon\gal$ is the mass ratio between the MBH and the host  
galaxy and $\epsilon\g$ the ratio between the mass accreted by the MBH  
in a MM event and the total gas mass available, both in the  
supply-limited accretion regime. $\tau\acc$ is the typical timescale  
of accretion onto the MBH (essentially the quasar duty cycle). $M\min$  
and $M\max$ are the lower and upper masses of haloes likely to harbour  
quasars. Finally, $\nu$ is the fraction of MMs giving rise to the  
reactivation of one observed quasar (in the case of the integrated  
quasar light there is no value for $\nu$ since both the predictions  
and observations are normalised at $z=2$). Errors correspond to  
$1\sigma$ intervals, calculated in the standard way for a  
multiparametric fit by $\chi^2$ minimization. The lack of errors in  
the second column is due to the poor fit obtained in this case, which  
makes such an estimation meaningless.}  
  
\label{tabres}  
\begin{center}  
\begin{tabular}{lllll}  
\hline  
\hline  
                        &~~~~~~2QZ &~~~~~~~~2QZ &~~~high-$z$ &~~~~~$\rho_L$\\  
                        & $0.3<z<2.3$                  &   $1.27<z<2.3$        & $3<z<4.7$     & $0<z<6$\\  
\hline  
\hline  
$z\refe$                        & $~~~~~1.92$  & $~~2.01^{+0.02}_{-0.02}$  &  
$~4.6^{+\infty}_{-4.6}$    & $~1.97^{+0.04}_{-0.05}$\\  
$\epsilon\gal~(\times 10^{-3})$ & $~~~~~7.91$ &$~~2.45^{+0.02}_{-0.05}$ &  
$~6.6^{+6.7}_{-2.6}$ &  $~2.3^{+\infty}_{-2.3}$\\  
$\epsilon\g~(\times 10^{-2})$   & $~~~~~5.08$  & $~~1.54^{+0.18}_{-0.18}$ &  
$~11.0^{+\infty}_{-10.5}$  & ~$8.5^{+\infty}_{-8.5}$\\  
$\tau\acc$  (Gyr)               & $~~~~~0.05$  & $~~0.18^{+0.01}_{-0.02}$  &  
$~0.001^{+0.005}_{-0.001}$ & $~0.02^{+0.06}_{-0.02}$\\  
$k$                             & $~~~~~9.21$  & $~~2.36^{+0.10}_{-0.09}$  &  
$~1.0^{+4.1}_{-0.5}$    & $~2.01^{+2.20}_{-0.17}$\\  
$\log M\min$ (M$_\odot$)        & $~~~~~10.72$ & $~~12.47^{+0.01}_{-0.01}$  &  
$~11.4^{+0.2}_{-\infty}$   &$~10.0^{+0.7}_{-1.2}$\\  
$\log M\max $ (M$_\odot$)       & $~~~~~12.88$ & $~~13.72^{+0.09}_{-0.04}$ &  
$~14.6^{+\infty}_{-2.0}$  &$~13.50^{+\infty}_{-0.6}$\\  
$\nu $                          & $~~~~~0.02$ & $~~0.11^{+0.01}_{-0.01}$ &  
$~0.05^{+0.07}_{-0.03}$  &  \,\,\,~~~~--  \\  
\hline  
\hline  
\end{tabular}  
\end{center}  
\end{table*}  
  
The different samples of quasars studied with redshifts ranging  
from 1.27 to 6, lead to values of the parameters that do not differ  
significantly. The only relevant exception is $\tau\acc$, the typical  
timescale of accretion onto the MBH, which shows a clear trend to  
diminish with increasing $z$ as actually expected. It is worth  
emphasizing that the values found for all the parameters are very  
reasonable. In particular, those of $\tau\acc$, ranging from $10^6$ to  
$2\times 10^8$ yr according to $z$, are consistent with the most  
recent estimates (\citealt{yu02, schir03}). The ratio  
$\epsilon\gal\equiv M\BH/M\gal$ is found to be $\simeq 2.5 \times  
10^{-3}$, in agreement with observations \citep{fermer00,gebhardt00}.  
The fraction $\epsilon\g$ of hot gas that ends up in the central MBH  
after a MM in the supply-limited accretion regime is of a few percent.  
The value $\sim 2$ of the parameter $k$ determining the accretion rate  
in the Eddington-limited regime is just what one would expect from \eq  
(\ref{ldet}) for a quasar of luminosity $L_8$, representative of a MBH  
of $10^8$ M$_{\odot}$, and a typical accretion rate of $\dot M_8$  
during a duty cycle of $10^8$ yr. On the other hand, according to  
\cite{cavitt00}, the transition between the Eddington-limited and the  
supply-limited accretion regimes is expected to occur at the typical  
formation of groups, at $z$ equal to 2.5$-$3.0, where MMs of haloes of  
the galactic scale become rarer and the accretion onto MBHs  
progressively declines. In our model, such a transition takes place  
when the gas fraction in baryons becomes so low that the product  
$\epsilon\g M\g$ falls below $k M\BH$, thus becoming the limiting  
factor. This occurs at the epoch where $M\gal/M\g \simeq  
\epsilon\g/(k\epsilon\gal)$, with the latter ratio taking a value of  
order unity, which is satisfied at $z\refe \sim 2$ not far from  
the redshifts preferred by \citet{cavitt00}. The hosting halo masses  
range from $\sim 10^{11}$ M$_{\odot}$ to $\sim 10^{14}$ M$_{\odot}$,  
implying that quasars would be found within central galaxies of haloes  
from the galactic scale to the scale of galaxy groups, the structures  
that are currently believed to be the harbouring environments of the  
brightest quasars (\citealt{jager01}). Finally, the inferred fraction  
of observed quasars per halo MM is found to be $\nu\simeq$ 5--10\%,  
again pretty close to the expectations.  
  
All these results, \ie the good behaviour of the predicted QLFs at  
$z>1$ and the reasonable best-fitting values of the parameters, give  
strong support to the proposed connection between halo MMs and AGN  
lightning through the triggered simultaneous feeding of the spheroid  
of the main galaxy and its central MBH. The fact that our model does  
not match the observed evolution of QLFs at $z<1$ is also not  
unexpected in this scheme because of the failure for late times of  
that central hypothesis. The two complementary mechanisms connecting  
halo MMs and the reactivation of AGN imply the parallel growth of the  
host galaxy (by cooling flows and the merger of the main galaxies of  
the progenitor haloes). Consequently, they can be expected to operate  
as far as haloes are typically of the galactic scale, {\it when  
galaxies are growing}. When haloes reach the scale of galaxy groups,  
galaxies within them no longer grow in size, but just in  
number. Therefore, none of the above mentioned processes may then be  
effective. At such late epochs, there may still be radial flows of gas  
into MBHs lying at the centre of galaxies, but they should be  
triggered by other mechanisms ---such as internal instabilities of  
galaxies or tidal interactions among galaxies within groups and  
clusters--- uncorrelated with halo MMs.  
  
These results also indicate the necessity of dealing more accurately  
with the physics of baryons, that is, with the processes governing the  
evolution of the hot gas trapped in haloes, as well as with the growth  
of galaxies and their interactions. In other words, a QLF model  
spanning from high to small redshifts should necessarily be coupled to  
a detailed model of galaxy formation and evolution as in the recent  
work by \citet{menci03}. There are other simplifications in our model  
that might be also improved in future works, although they do not seem  
to play a crucial role on our main conclusions. There is, for  
instance, no consideration of the detailed physics at the vicinity of  
the MBH and our radiation model does not account for relativistic  
effects. Detailed study of the physics around the MBH could provide,  
for example, a natural explanation to the decrease of the number of  
bright quasars (\citealt{nitta99}).  Likewise, the use of an isotropic  
model to describe highly anisotropic radiation fields around quasars  
is likely to introduce different kinds of biases. The first one, due  
to the requirement that the line-of-sight lies in the emission cone,  
is of purely statistical nature and should be reasonably accounted for  
by the normalisation of the MM rate. A second source of bias arises  
from the relation between the bolometric luminosity and the MBH mass,  
through the Eddington efficiency, which assumes isotropy of the  
emission. This approximation, which cannot be corrected by a shift of  
the luminosity function along the magnitude axis, results in erroneous  
masses associated with the observed luminosities and, therefore, in  
erroneous merger rates.  Finally, one must not forget that the  
comparison with observational data usually takes into account only  
statistical errors (very small in such large samples as the 2QZ).  
Systematic effects, however, like the variations of the ability of  
redshift determination depending on the emission lines present inside  
the observable wavelenght range, are likely to have a major  
contribution (\citealt{croom01a}).  
  
\vspace{0.75cm} \par\noindent    
{\bf ACKNOWLEDGMENTS} \par   
   
\noindent We would like to thank the people of the Toulouse  
Extragalactic Team as well as Hugo Capelato, Gary Mamon, Joseph Tapia  
and many others for help and stimulating discussions, and Brian Boyle  
for kindly providing us with the latest QLF of the 2QZ survey. G.M.\  
acknowledges funding from \'Egide and hospitality from IAG-USP and  
INPE, while E.S.S.\ acknowledges funding from the Observatoire  
Midi-Pyr\'en\'ees and the hospitality of its staff.

\begin{appendix}   
\section{KINETIC APPROACH FOR THE HALO MM RATE}\label{kinetic}   
  
\begin{figure}  
\centerline{\psfig{file=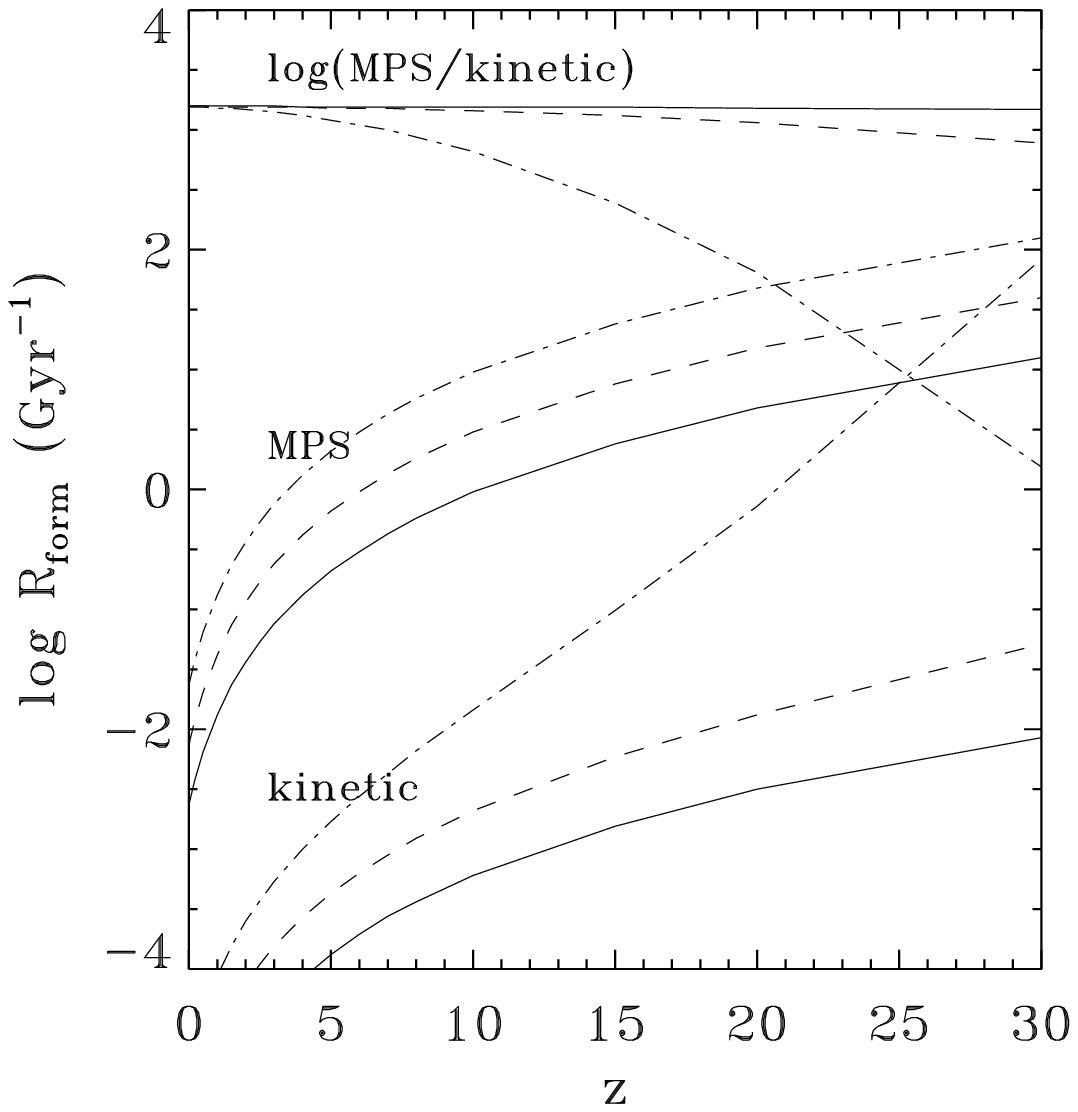,height=9cm,width=9cm}\quad}  
\caption{Logarithm of the halo formation rate vs.\ redshift in the  
Einstein-de Sitter cosmology with an $n=0$ power-law spectrum. Two  
sets of curves have been inferred from the MPS model and the kinetic  
theory (the latter with $K=1$ in \eq [\ref{K}]), for hosting halo  
masses $M\H=10^{12}$ M$_\odot$ (solid lines), $M\H=10^{13}$ M$_\odot$,  
(dotted lines) and $10^{14}$ M$_\odot$ (dashed lines). On the top,  
logarithm of the corresponding ratio, which is constant and equal to  
$1550$ within 2\% out to $z=20$ for $M\H=10^{12}$ M$_\odot$ and within  
12\% out to $z=3$ for $M\H=10^{14}$ M$_\odot$.}  
  
\label{MRT}      
\end{figure}      
  
In the simplest case of an $n=0$, power-law power spectrum of density   
fluctuations (where there is no correlation among haloes), the merger   
rate is simply given, in the kinetic theory, by the usual expression   
\begin{eqnarray}          
\nnb R(M\H,t)\,\der M\H = \qquad\qquad  
\qquad\qquad \qquad\qquad \qquad\qquad \\          
\int\p f(M\H,t)\,\der M\H\,f(M\ph,t)\,\der M\p\,   
\langle v\,S(v)\rangle\p\,(1+z)^3\,,  
\label{krmm}          
\end{eqnarray}          
with $M\p$ and $M\ph\equiv M\H-M\p$ the masses of the primary and  
secondary progenitor haloes, respectively, $\langle v\,S(v)\rangle\p$  
the product of the relative velocity between the two progenitor haloes  
and their cross section for merger averaged over those velocities, and  
$(1+z)^3$ the factor that brings the latter product, expressed in natural  
units, to comoving units as the quantities entering in this  
expression. A reasonable approximation to adopt is  
\begin{eqnarray}          
\langle v\,S(v)\rangle\p = K\, \pi r\p^2\,\sigma\p\label{K} \,,  
\end{eqnarray}   
with $r\p$ and $\sigma\p$ the gravitational radius and internal 3-D  
velocity dispersion, respectively, of the primary progenitor, which can  
be estimated from their scaling relations with $M\p$. Indeed, provided  
all relaxed haloes have a common density contrast $\delta$, one has  
within the radius $r$ that $M = \delta\,r^3\,\rho =  
\delta\,r^3\,\rho_0\, \left(1+z \right)^3$, with $\rho$ the mean cosmic  
density at redshift $z$, leading to  
  
\begin{eqnarray}         
r\p \propto M\p^{1/3} (1+z)^{-1}\,.  
\end{eqnarray}          
On the other hand, from the virial relation $2T+U=0$ between the halo     
kinetic energy $T\propto M \sigma^2$ and its potential energy     
$U\propto GM^2/r \propto GM^{5/3}$, the velocity dispersion takes the     
form       
\begin{eqnarray}          
\sigma\p \propto M\p^{1/3} (1+z)^{1/2}\,.\label{sp}  
\end{eqnarray}    
Finally, the two preceding proportionalities can be calibrated from     
the results of \citet{wilson01} on galaxy-galaxy lensing which    
give the following parameters for a halo harbouring an $L_\ast$      
elliptical      
\begin{eqnarray}          
M_\ast &=& 1.3 \times 10^{12}\;h^{-1}\;{\rm M_{\odot}} \nnb\\          
r_\ast &=& 100 \; (50 \;\;{\rm to}\;\; 200)\;h^{-1}\;{\rm kpc}\nnb\\     
\sigma_\ast &=& 255\;(240\;\;{\rm to}\;\;270)\;{\rm km~s^{-1}}\,.\nnb   
\end{eqnarray}          
This leads to     
\begin{eqnarray}       
r\p &=& r_\ast\left(\frac{M\p}{M_\ast}\right)^{1/3}(1+z)^{-1}\label{rp}\\   
\sigma\p &=& \sigma_\ast \left(\frac{M\p}{M_\ast}\right)^{1/3}      
(1+z)^{1/2}\,.\label{sigmap}         
\end{eqnarray}    
  
With these approximations, the ratio between the formation rates  
$R\form$ inferred from the MPS model and the kinetic theory (eqs.  
[\ref{mmr}] and [\ref{krmm}], respectively) in the Einstein-de Sitter  
cosmology is constant and equal to unity for $K=1550$ (see \fig  
\ref{MRT}), except for large masses relative to the typical mass for  
collapse where the kinetic approach is no longer valid owing to the  
rarity of haloes. In this simple scale-free, $n=0$ power-law  
cosmology the behaviour of the MM rate as the product of two PS mass  
functions for moderate and small $M\H$ can be seen from \eq  
(\ref{epsmr}) by taking into account the corresponding explicit form of  
$\sigma(M\H)$. Although this is not so obvious in the general case, the  
fact that one always finds very similar results to those plotted in  
\fig \ref{MRT} allows extending that conclusion to any arbitrary  
cosmology.

\section{MBH AND HALO PROPERTIES}\label{team}  
  
\subsection{Growth of haloes}  
  
\begin{figure}  
\centerline{\psfig{file=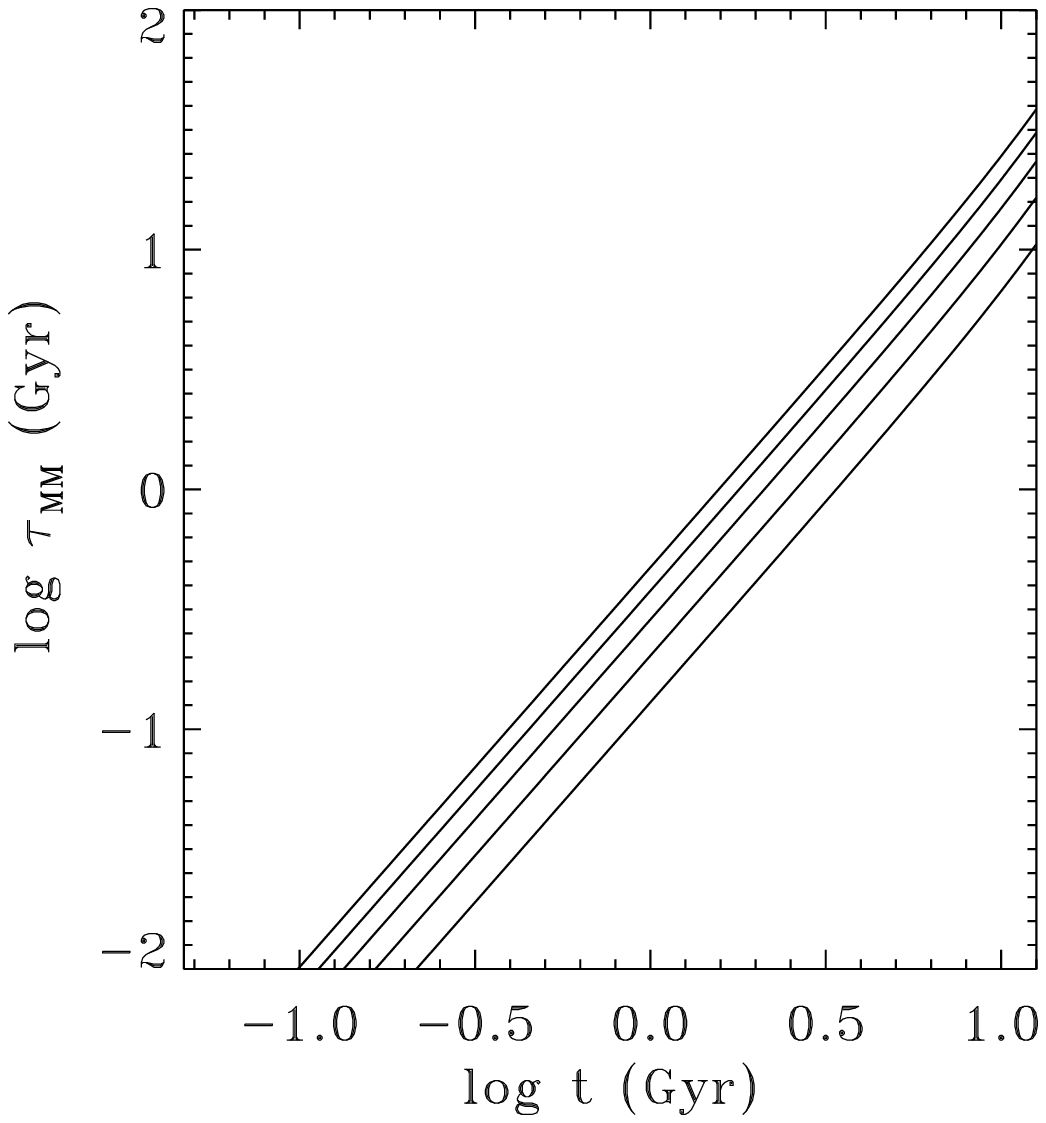,height=9cm,width=9cm}\quad}    
\caption{Typical time interval between MMs vs.\ cosmic time for  
halo masses ranging from $10^{11}$ to $10^{15}$ M$_\odot$ (top to  
bottom).}  
\label{MR}    
\end{figure}    
  
As shown in \fig (\ref{MR}), the time dependence of $\tau\MM(M\H,t)$  
for a fixed mass $M\H$ is very close to a power-law and the mass  
dependence for a fixed $t$ is not far from this simple law, too. Thus,  
\begin{equation}  
\tau\MM(M\H,t) = \tau_*\,  
\left(\frac{t}{t_*}\right)^{a+1}\,\left(\frac{M_H}{M_*}\right)^{-b} \,,  
\label{tauMM}  
\end{equation}  
with $\tau_*$, $a$ and $b$ equal to 0.73, 0.71 and 0.15, respectively,  
$M_* = 10^{11}$ M$_\odot$ and $t_* = 1$ Gyr. By bringing this general  
dependence into \eq (\ref{iv}), we are led to  
\begin{equation}  
\frac{\der M\H}{\der t} \approx \frac{M_*^b}  
{\tau_*}\,\,\left(\frac{t_*}{t}\right)^{a+1}\,\left[M\H(t\cos)\right]^{1-b}\,,  
\label{B2}  
\end{equation}  
which easily integrates to  
\begin{equation}  
\left[\frac{M_*}{M\H(t\cos)}\right]^b = \left(\frac{M_*}{M\refe}\right)^b -  
B\left[\left(\frac{t_*}{t\refe}\right)^a -  
\left(\frac{t_*}{t\cos}\right)^a \right]\,, \label{MH}   
\end{equation}  
with $B=(bt_*)/(a\tau_*)$. The set of functions $\{M\H(t\cos)\}$ is a  
one-parameter family, with each member specified by its value $M\refe$  
at some reference time $t\refe$.  
  
All halo masses tend to zero as cosmic time tends to zero by virtue of  
\eq (\ref{MH}). On the other hand, when time tends to infinity, the  
second member of \eq (\ref{MH}) tends to  
\begin{equation}  
\left(\frac{M_*}{M_{\infty}}\right)^b = \left(\frac{M_*}{M\refe}\right)^b-  
B\,\left(\frac{t_*}{t\refe}\right)^a\,.  
\label{MHinfinite}   
\end{equation}  
The requirement that both members of \eq (\ref{MH}) are positive  
at all redshifts imposes a condition on $M\refe$, which  
cannot be arbitrarily large but has to be smaller than  
\begin{eqnarray}  
M\top \equiv M_* \left[B^{-1}\left( \frac{t\refe}{t_*} \right)^a   
\right]^{1/b}\hskip-3pt  
\simeq 2.7 \times 10^{14} \left(\frac{t\refe}{t_*}\right)^{4.7}  
\;\hskip-3pt {\rm M}_{\odot}\,.\label{infini}  
\end{eqnarray}  
Otherwise, any halo of mass $M\refe>M\top$ at time $t\refe$ would  
reach an infinite mass at some finite time. Conversely, given a fixed  
mass $M\refe=M\H$, the corresponding time of reference $t\refe$ must  
be larger than the value $t\top$ satisfying  
\begin{equation}  
t\top = t_* \left[B\, \left( \frac{M\H}{M_*}\right)^b \right]^{1/a}\,.  
\end{equation}  
  
\subsection{Growth of Massive Black Holes}\label{evolgas}   
  
\subsubsection{Eddington-limited accretion regime}\label{ss}  
  
According to the assumption (\ref{k}), equation (\ref{v}) writes  
\begin{equation}   
\frac{\der M\BH}{\der t} \approx k\,\frac{M\BH[M\H(t),t\cos]}  
{\tau\MM[M\H(t),t]}\label{vibis}  
\end{equation}    
and, taking into account \eq (\ref{iv}), we get  
\begin{equation}  
\frac{\der M\BH}{M\BH}\approx k\,\frac{\der M\H}{M\H}\,.  
\end{equation}    
This leads, through \eq (\ref{B2}), to   
\begin{equation}   
M\BH(t)={M\BH}\refe\,\left[\frac{M\H(t)}{M\refe}\right]^k  
\label{Ereg}  
\end{equation}  
and, from \eq (\ref{k}), we also get the solution $\Delta M\BH(t)$. In   
\eq (\ref{Ereg}), ${M\BH}\refe$ stands for ${M\BH}(M\refe,t\refe)$, the  
boundary condition specified below.  
  
For any pair of physically acceptable values $({M\H}_i,t_i)$ (\ie with  
corresponding value of $M\refe$ at $t\refe$, determined through \eq  
[\ref{MH}] for $M\H(t_i)={M\H}_i$, satisfying the constraint  
[\ref{infini}]), the value $M\BH({M\H}_i,t_i)$ is simply equal to the  
solution (\ref{Ereg}) for $t=t_i$.  
  
\subsubsection{Supply-limited accretion regime}  
  
Equations (\ref{v}) and (\ref{iii}) and the time derivative of  
\eq (\ref{ii}) lead to  
\begin{equation}   
\frac{\der M\gal}{\der t}\approx\frac{\epsilon\g}{\epsilon\gal}\;   
\frac{M\g[M\H(t),t\cos]}  
{\tau\MM[M\H(t),t]}\,,\label{vi}  
\end{equation}    
while \eq (\ref{iv}) and the definition of $M\bar$ leads to  
\begin{equation}  
\label{equadiff2} \frac{\der M\bar}{\der t} \approx \frac{ M\bar[M\H(t)]}  
{\tau\MM[M\H(t),t]}\,.  
\end{equation}  
By defining the baryonic fraction in hot gas, $\gamma(M\H,t\cos)\equiv  
M\g(M\H,t\cos)/M\bar(M\H,t\cos)$, \eqs (\ref{vi}) and  
(\ref{equadiff2}) lead to  
\begin{equation}  
\label{equadiff3} \tau\MM[M\H(t),t] \;\frac{\der\gamma}{\der t}+  
\left(1+\frac{\epsilon\g}{\epsilon\gal}\right)  
\;\gamma(t) \approx 1\,,  
\end{equation}  
with trivial peculiar solution $\gamma_0$ equal to  
$\left(1+\epsilon/\epsilon\gal\right)^{-1}$. Then, by replacing \eqs  
(\ref{tauMM}) and (\ref{MH}) into \eq (\ref{equadiff3}) and solving  
the differential equation, one gets  
\begin{eqnarray}  
\gamma(t)=\gamma_0+(\gamma\refe-\gamma_0)\,  
\left[\frac{M\refe}{M\H(t)}\right]^{1+\frac{\epsilon\g}{\epsilon\gal}}\,,  
\label{gammasol}  
\end{eqnarray}  
with $\gamma\refe$ the value of $\gamma$ at $t\refe$.  
  
As in \ref{ss}, for any pair of physically acceptable values  
$({M\H}_i,t_i)$, the value $\gamma({M\H}_i,t_i)$ is given by the  
solution (\ref{gammasol}) for $t=t_i$. The boundary condition  
$\gamma\refe = \gamma(M\refe,t\refe)$ necessary to find this latter  
solution is actually unknown, as it requires to follow in detail the  
physics of baryons trapped in haloes of all masses since the dark  
ages. However, since we expect the change in accretion regimes from  
Eddington to supply-limited to take place when the baryonic  
fraction in hot gas becomes smaller than some threshold $\gamma\refe$,  
we can assume that the regime changes at some time $t\refe$ (or  
redshift $z\refe$), for which all haloes have similar values of  
$\gamma$ equal to that threshold, i.e.,  
$\gamma(M\refe,t\refe)=\gamma\refe$.  
  
Therefore, for any given couple of values of $z\refe$ and  
$\gamma\refe$ to be determined, one can infer the function  
$\gamma(M\H,t)$ and, from it and \eqs (\ref{i})--(\ref{v}), all  
quantities of interest; first  
\begin{eqnarray}  
M\g(M\H,t\cos)   =  \gamma(M\H,t)\,f\bar\,M\H\qquad\quad\label{1} \\  
M\gal(M\H,t\cos)  = [1-\gamma(M\H,t)]\,f\bar\,M\H\label{2}\,,  
\end{eqnarray}  
and then  
\begin{eqnarray}   
M\BH(M\H,t\cos)  =  \epsilon\gal \, M\gal(M\H,t\cos)\quad\label{1bis}\\  
\Delta M\BH(M\H,t\cos)  =  \epsilon\g \, M\g(M\H,t\cos)\,.\label{2bis}  
\end{eqnarray}  
The values of these two latter functions at $z\refe$ fix the boundary  
condition for the solution in the Eddington-limited regime, given by  
equations (\ref{Ereg}) and (\ref{k}). Let us finally note that by  
dividing equation (\ref{2bis}) by equation (\ref{1bis}) at $z\refe$,  
where the solutions $M\BH(M\H,t)$ and $\Delta M\BH(M\H,t)$ in both regimes  
match and, taking into account equations (\ref{1}) and (\ref{2}), we  
obtain  
\begin{equation}  
\gamma\refe=\frac{k}{k+\frac{\epsilon\g}{\epsilon\gal}}\,.  
\end{equation}  
Thus, instead of $z\refe$ and $\gamma\refe$, we can use $z\refe$ and  
$k$ (for a given value of $\epsilon\g/\epsilon\gal$) as the  
boundary conditions fixing the evolving properties of MBHs.  
\end{appendix}   
   
\end{document}